\documentclass[a4paper,fleqn,usenatbib]{mnras}

\usepackage{newtxtext,newtxmath}

\usepackage[T1]{fontenc}
\usepackage{ae,aecompl}
\usepackage{hyperref}

\usepackage{graphicx}	
\usepackage{amsmath}	
\usepackage{amssymb}	

\title[Detecting quasar halo heating with ALMA]{Detecting the halo heating from AGN feedback with ALMA}

\author[S Brownson et al.]{
S. Brownson$^{1,2}$\thanks{E-mail: sbb33@cam.ac.uk}, R. Maiolino$^{1,2}$, M. Tazzari$^{3}$, S. Carniani$^{4}$, N. Henden$^{3}$
\\
$^{1}$Kavli Institute for Cosmology, University of Cambridge, Madingley Road, Cambridge CB3 0HA, UK\\
$^{2}$Cavendish Laboratory, University of Cambridge, 19 J. J. Thomson Ave., Cambridge CB3 0HE, UK\\
$^{3}$Institute of Astronomy, University of Cambridge, Madingley Road, Cambridge CB3 0HA, UK\\
$^{4}$Scuola Normale Superiore, Piazza dei Cavalieri 7, I-56126 Pisa, Italy\\
}

\date{Accepted XXX. Received YYY; in original form ZZZ}

\pubyear{2018}

\begin{document}
\label{firstpage}
\pagerange{\pageref{firstpage}--\pageref{lastpage}}
\maketitle

\begin{abstract}
The Sunyaev-Zel'dovich (SZ) effect can potentially be used to investigate  the heating of the circumgalactic medium and subsequent suppression of cold gas accretion onto
the host galaxy caused by quasar feedback. We use a deep ALMA observation of HE0515-4414 in band 4, the most luminous
quasar  known at the peak of cosmic star formation (z=1.7), to search for the SZ signal tracing the
heating of the galaxy's halo. ALMA's sensitivity to a broad range of spatial scales enables us to disentangle
emitting compact sources from the negative, extended SZ signal.
We obtain a marginal S-Z detection ($\sim$3.3$\sigma$) on scales of about 300 kpc (30--40 arcsec),
at the 0.2~mJy level, 0.5~mJy after applying a correction factor for primary beam attenuation and flux that is resolved out by the array.
We show that our result is consistent with a simulated ALMA observation of a similar
quasar in the \textsc{fable} cosmological simulations. We emphasise that detecting an SZ signal is more easily achieved in the visibility plane
than in the (inferred) images.
We also confirm a marginal detection ($3.2\sigma$) of a potential SZ dip on smaller scales ($<$100~kpc)
already claimed by other authors, possibly highlighting the
complex structure of the halo heating. 
Finally, we use SZ maps from the \textsc{fable} cosmological simulations, convolved with ALMA simulations,
to illustrate that band 3 observations
are much more effective in detecting the SZ signal with higher significance, and discuss the optimal observing
strategy.

\end{abstract}




\section{Introduction}
Feedback mechanisms, in which AGN inject large
amounts of energy into their host galaxies and the circumgalactic medium, are commonly invoked to
explain various properties of galaxies and scaling relations between supermassive black holes  (SMBHs) and their host galaxies
\citep{Fabian2012,Silk2012,King2015}.
Interestingly, most galaxies containing AGN lie in the green valley - a region in the colour-magnitude diagram
separating the blue and red populations \citep{Fabian2012}. In fact, \citet{Schawinski2010}  have exploited the
location and fraction of AGN in the colour-magnitude diagram, particularly in the green valley, as a proxy for AGN
duty cycle to characterise the importance of AGN in the evolution of different galaxy types. The universal
shutdown of star formation, which peaked  at ${z\approx2}$ \citep{Madau2014}, is thus often attributed to the
influence of AGN, at least in massive galaxies.

Feedback mechanisms are also capable of producing a number of the observed correlations between the mass of BHs and their host galaxies, such as the relation  between the BH mass (${M}_{BH}$) and the stellar velocity dispersion, as well as with the galaxy's bulge mass (${M}_{Bulge}$) \citep{Kormendy2013, King2015}.

A simple picture in which the stellar mass function follows the form of the halo mass function predicts the
  existence of too many galaxies at both the low and high mass ends. The steep decline in the stellar mass function
  for high mass galaxies can be explained in terms of AGN feedback. High mass galaxies typically have more powerful AGN that can
  exert more influence on their host's stellar population, suppressing star formation \citep{Silk2012}. Cosmological simulations have
  successfully reproduced this downturn in the stellar mass function through the inclusion of AGN feedback
  \citep[e.g.][]{Croton2006,Torrey2014}. The decline in the stellar mass function at the low mass end is typically attributed to star
  formation and stellar feedback \citep[e.g.][]{Chisholm2017}.

The AGN feedback paradigm is further supported by energetic considerations. Assuming fairly modest accretion efficiencies ($\sim$10 per cent),  the energy emitted by a BH over the course of its growth is shown to be $\sim$100 times larger than a galaxy's binding energy \citep{Fabian2012}. Only a small fraction of the BH's energy is needed to couple to its host for the AGN to have a profound effect on the evolution of the host galaxy, such as the suppression of star formation. Yet we require a framework that couples the energy radiated by AGN to their environments in order to fully describe the coevolution of BHs and their galaxies. Two such frameworks are commonly invoked: kinetic and thermal \citep{Husemann2018}. 

The kinetic (ejective)
framework, in which AGN drive massive, extended (up to $\sim$30~kpc) outflows into their host galaxies, has
traditionally been viewed as the primary mechanism for quenching, ejecting gas form galaxies and depriving them of
fuel for star formation \citep[e.g.][]{Fabian2012,King2015,Springel2005}.  These multi-phase outflows are observed both locally and at high
redshift \citep[e.g.][]{Maiolino2012, Cicone2014, Cicone2015, Fluetsch2019}.
The primary driving mechanism responsible for outflows
has been greatly debated, with some advocating an energy-conserving blast-wave scenario \citep[e.g.][]{King2015}
and others favouring direct radiation pressure \citep{Ishibashi2018}. Efficient,
energy-driven winds were supported by early
observations \citep{Feruglio2010,Cicone2014, Cicone2015, Fiore2017}, but recent works \citep{Feruglio2017,Bischetti2018,Fluetsch2019}
have suggested that the ejective mode of outflows may not be as effective as initially thought and that
they may be either energy-driven with modest coupling with the galaxy ISM as also suggested
by some models \citep{Costa2014,Gabor2014,Roos2015,Hartwig2018} or radiation-pressure driven.
These results show that AGN-driven outflows may not be effective at completely ejecting gas and raise doubts about their quenching effect.  

\citet{Peng2015} and \citet{Trussler2019} have compared the stellar metallicities of star forming and passive galaxies
to argue that \textit{starvation}, rather than kinetic gas removal, is the primary mechanism for quenching star
formation. This alludes to the second feedback framework: thermal coupling. Here, thermal energy is injected on large
circumgalactic scales, significantly reducing the cooling rate of the halo gas
\citep[e.g.][]{Croton2006,Dekel2006,Cattaneo2009,Ciotti2010,Weinberger2017, Nelson2019}. The injection of energy happens mainly through radio jets in AGNs with low
accretion rate in early models. This mode is therefore often also refereed to as the ``radio mode''. However, more recent models and
simulations have shown that shocking through outflows may be another important cause of circumgalactic
gas heating. These AGN heating modes probably combine with the gravitational shock heating to keep the halo hot
in massive systems. Regardless of the model, the injection of thermal energy prevents
further accretion of pristine gas, thereby starving the galaxy of the fuel needed for continued star formation. Rather
than the rapid shutdown in star formation implied by the kinetic/ejective mode,
the thermal mode suggests a delayed (or ``preventive'') form of feedback.

Thermal feedback has been successfully incorporated into simulations, but a suitable observational probe is yet to be
found. X-ray observations are commonly used to trace hot gas \citep{Fabian2012}.  Thermal bremsstrahlung, however, has
a quadratic density dependence, rendering X-ray observations ineffective as a probe for the large scale heating of
diffuse gas. While hot halos are commonly observed in the X-rays in nearby galaxy clusters, X-ray observations of the
circumgalactic medium of individual galaxies are much more difficult. Moreover, X-ray diffuse emission associated with
circumgalactic or intracluster gas suffers from a cosmological dimming as $\rm (1+z)^{-4}$, making detection
even more challenging at high redshift where the bulk of star formation quenching is expected to take place.

A second probe, the \textit{Sunyaev-Zel'dovich}  (SZ) effect, has attracted increasingly more attention. 
The SZ effect \citep{Sunyaev1972} is a secondary CMB anisotropy caused by the scattering of CMB
photons by high energy electrons.
The thermal SZ effect, the version relevant for
this work, is the inverse Compton scattering of CMB photons as they traverse hot ionised gas. A full mathematical
treatment of the SZ effect can be found in \citet{Birkinshaw1999}, a review paper. The energy of the radiation field
increases as energy is transferred from high velocity electrons in the plasma to the CMB photons. Given  conservation
of photon number, the average photon energy must increase, forcing a shift of the overall CMB SED towards higher
frequencies \citep{Sunyaev1972}. This energy boost leaves a frequency dependent signature in the SZ SED: a decrement
in brightness in the Rayleigh-Jeans region and an increment in the Wien region. The transition from decrement to
increment occurs at $\sim218~GHz$. The SZ effect should be observable both above and below this frequency. However, it is
difficult to uniquely associate any observed positive continuum emission with the SZ increment since it could be
easily confused with other common sources of continuum emission, such as the thermal emission from  dust grains in
galaxies.  Negative continuum emission, however, is not observed elsewhere in the universe and can be confidently
attributed to an SZ effect. Crucially, the magnitude of the SZ decrement is sensitive to the integrated pressure of the ionised gas and, therefore, to additional heating from AGN feedback. 
 
While the surface brightness of most sources of emission 
decreases with increasing redshift as ${(1+z)}^{-4}$, the observed brightness of the SZ effect is
independent of redshift (modulo the intrinsic redshift evolution of the heating processes).
This feature  makes the SZ effect an unbiased cosmic probe, allowing us to confidently search for and characterise feedback physics at all redshifts. 

SZ signals have been detected in many galaxy clusters. Attempts have been made
to detect an SZ signal associated with the halo of individual galaxies, and possibly associated with AGN/quasar heating, 
by stacking observations from single dish telescopes and space observatories. However, these attempts (which use
data with large beams on the sky) have been plagued by the difficulties of disentangling
any potential SZ signal from the host galaxies mm-IR emission. The positive emission associated with
individual galaxies and the negative SZ signal are distributed on vastly different angular scales.
Interferometric observations have exploited the range of spatial scales to remove contaminating sources for
mapping the SZ signals of galaxy clusters \citep{Jones1993}. In this work, we investigate the use of interferometers to detect the SZ signal from the halos of single galaxies.
 
We analyse an archival ALMA band 4 observation of the most powerful radio-quiet quasar
known (HE0515-4414) \citep{Reimers1998} at the epoch of the peak of cosmic star formation (z=1.7) with the
goal of exploring the detectability of the SZ signal associated with the putative hot halo of this system. This quasar has also been observed with a long exposure time with ALMA, making it a strong candidate for observing an SZ signal.
We also
use the results of cosmological simulations to explore the optimisation of future observing strategies
for detecting the SZ with ALMA. 

The paper is structured as follows. In Section \ref{ALMA observations and data reduction}, we describe the observation and subtraction of IR emission. We present the SZ emission in Section \ref{Results} and quantify the significance of the result. We put the result in context in Section \ref{Discussion} by comparing it with previous observations and the predictions from cosmological simulations. We then use these simulations to advise optimal observational setups for future observations. Finally, we summarise and conclude in Section \ref{Conclusion}. We assume a  $\Lambda$CDM cosmology  throughout, with H$_{0}$ = 70 km s${}^{-1}$ Mpc${}^{-1}$, ${\Omega}_{M}$ = 0.3 and ${\Omega}_{\Lambda}$ = 0.7.

\section{ALMA observations and data reduction}\label{ALMA observations and data reduction}
HE0515-4414 was observed during ALMA cycle 4 in band 4 using the 12m array with a total on-source time of 11.6~hours
(Project code: 2016.1.00309.S, PI M.Lacy; \citet{Lacy2018}). The quasar was observed  between January and March 2017 across 14 execution blocks (EBs) with the phase centred at  the location of the quasar (RA = 05:17:07.63, Dec = -44.10.55.5).  The array included $\sim$45 antennas with a minimum baseline length of 15m and maximum of  384m, producing a synthesised beam with size 3.2$\times$2.3~arcsec$^{2}$ and a corresponding spatial resolution of $\sim$20~kpc at $z\sim1.7$.

The spectral set-up has a total bandwidth of 8~GHz, with four spectral windows (SPWs) covering the band. The SPWs are centred at 133, 135, 145 and 147~GHz, each with 128 channels covering a 2~GHz band. The field of view (FoV) of the observation is thus $\sim$40~arcsec. 

The data were calibrated using the ALMA pipeline in CASA 4.7.2 \citep{Mcmullin2007}; J0519-4546 was used as
the phase calibrator, and J0538-4405 and Mars were used as flux calibrators. This paper considers
projected baseline lengths on the scale of the ALMA antenna diameters, 12m. The risk of shadowing is thus an
important consideration, particularly for observations tracking the source at low elevation. We note that the
pipeline flagged shadowed antennas with a tolerance level set to zero - i.e. the projected baseline lengths
were required to be larger than the sum of the radii of the antennas in each baseline. We double checked that 
shadowed data were flagged and discarded. The observations were proposed for with a required sensitivity of 4.7~$\mu$Jy~beam$^{-1}$ for the whole band.   The calibration yields an absolute flux uncertainty of $\sim$5 per cent at the observed frequency \footnote{See \href{https://almascience.nrao.edu/documents-and-tools/cycle4/alma-technical-handbook}{https://almascience.nrao.edu/documents-and-tools/cycle4/alma-technical-handbook} for details of flux calibration accuracy.}.

\subsection{Identification of emission sources}\label{Positive Emission}
The observation was proposed to target SZ continuum emission. The SPWs were thus chosen to be free
from any spectral line emission. Indeed, no lines were seen in any of the channels. Any potential SZ signal, however,
may be contaminated by positive galactic mm-infrared continuum emission, such as dust thermal, free-free and synchrotron emission.
HE0515-4414 also lies in the background of a $z=0.38$ galaxy group \citep{Bielby2016}. These sources are easily
detectable in such deep observations. The field therefore contains a number of emitting sources.

\begin{figure}
	\includegraphics[width=\columnwidth]{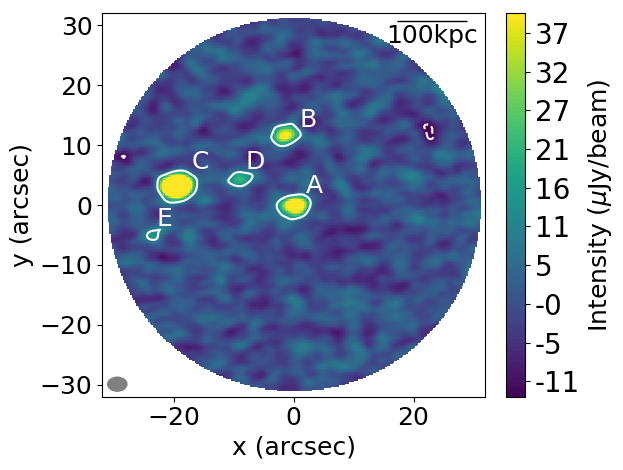}
    \caption{Clean, continuum image of HE0515-4414, the central source, and its environment - hosting four contaminating positively emitting sources. This map, and all other maps presented in this work, has no primary beam correction applied, and thus the rms noise level is uniform across the FOV. The emission is attenuated by a factor of five at the edges. The ellipse in the lower left corner denotes the synthesised beam size, and takes the same meaning for all future maps presented in this work. Contours denote the $\pm$3$\sigma$ levels, where the 1$\sigma$ level is at 3.8~$\mu$Jy~beam$^{-1}$.}
    \label{fig:PositiveContinuumEmission}
\end{figure}

We channel averaged the data across the 128 channels in each of the four SPWs and time averaged the data into  30 second intervals. This decreases the size of the data set significantly, by a factor of $\sim$2000, without significant loss in uv-coverage.
 Although we will focus on the analysis in the visibility plane, as discussed
in the following sections,  
we have also produced an image of the continuum emission using the CASA task \textit{tclean} with the the H\"{o}gbom deconvolving algorithm \citep{Hogbom1974}. A natural weighting was employed to maximise the
sensitivity, albeit at the expense of reduced angular resolution.  Such a compromise is necessary and worthwhile in  low signal-to-noise ratio
(SNR) work. We also chose to use a single Taylor coefficient in the spectral model due to the low SNR and relatively low fractional bandwidth of 10 per cent\footnote{See  \href{https://casa.nrao.edu/docs/taskref/tclean-task.html}{https://casa.nrao.edu/docs/taskref/tclean-task.html} for a discussion of suitable tclean parameters.}. The positive continuum emission is shown in Fig.
\ref{fig:PositiveContinuumEmission}, where an RMS noise level of 3.8~$\mu$Jy~beam$^{-1}$ is reached.  Five sources are observed with SNR
larger than three, with the central emission coming from HE0515-4414. Throughout this paper, we refer to the sources as A, B, C, D and E, as labelled in Fig. \ref{fig:PositiveContinuumEmission}. From Fig. \ref{fig:PositiveContinuumEmission} alone, it is clear that sources A, B and C  are slightly resolved and should not be treated as point-like. 

\subsection{Subtraction of emission sources}
Detecting the negative signal from an SZ  field is complicated by the positive continuum emission of the sources in the field. Thus the positive emission
must be properly modelled and/or removed before any SZ emission can be observed. We discuss the principles and methodology behind the source subtraction procedure in this section. This assumes an understanding of interferometric theory. We refer the reader to \citet{Thompson2017} for a full  discussion of the assumed concepts.

\subsubsection{Modelling in the Fourier Plane}
Fitting a model to interferometric data is better done in the Fourier plane that in the image plane.
Indeed, the first interferometric observation of an SZ signal in galaxy clusters fitted models
directly to the visibility data   \citep{Jones1993}.  An accurate representation of the sky brightness is
required before model parameters can be inferred in the image plane. Yet, incomplete sampling of the $(u,v)$
plane implies that the resulting synthesised image is not unique. As a result, estimating uncertainties on quantities inferred from the images is not straightforward.  Working in the visibility plane (i.e. fitting the visibilities) is a straightforward operation  on the interferometric measurements which allows a direct computation of uncertainties. One can fit a model sky brightness distribution by finding its Fourier transform and fitting model visibilities,
$Vis_\mathrm{mod}(u,v)$, with the same $(u,v)$ coordinates as the observed data set, $Vis_\mathrm{obs}(u,v)$. Furthermore, the noise is better understood
in the visibility plane. Each visibility measurement has  simple Gaussian noise associated with it. On the contrary,
the Fourier transform in the measurement equation \citep{Thompson2017} produces correlated noise in the image plane, thereby further complicating any fitting procedure in the image plane. 

\begin{figure}
	\includegraphics[width=\columnwidth]{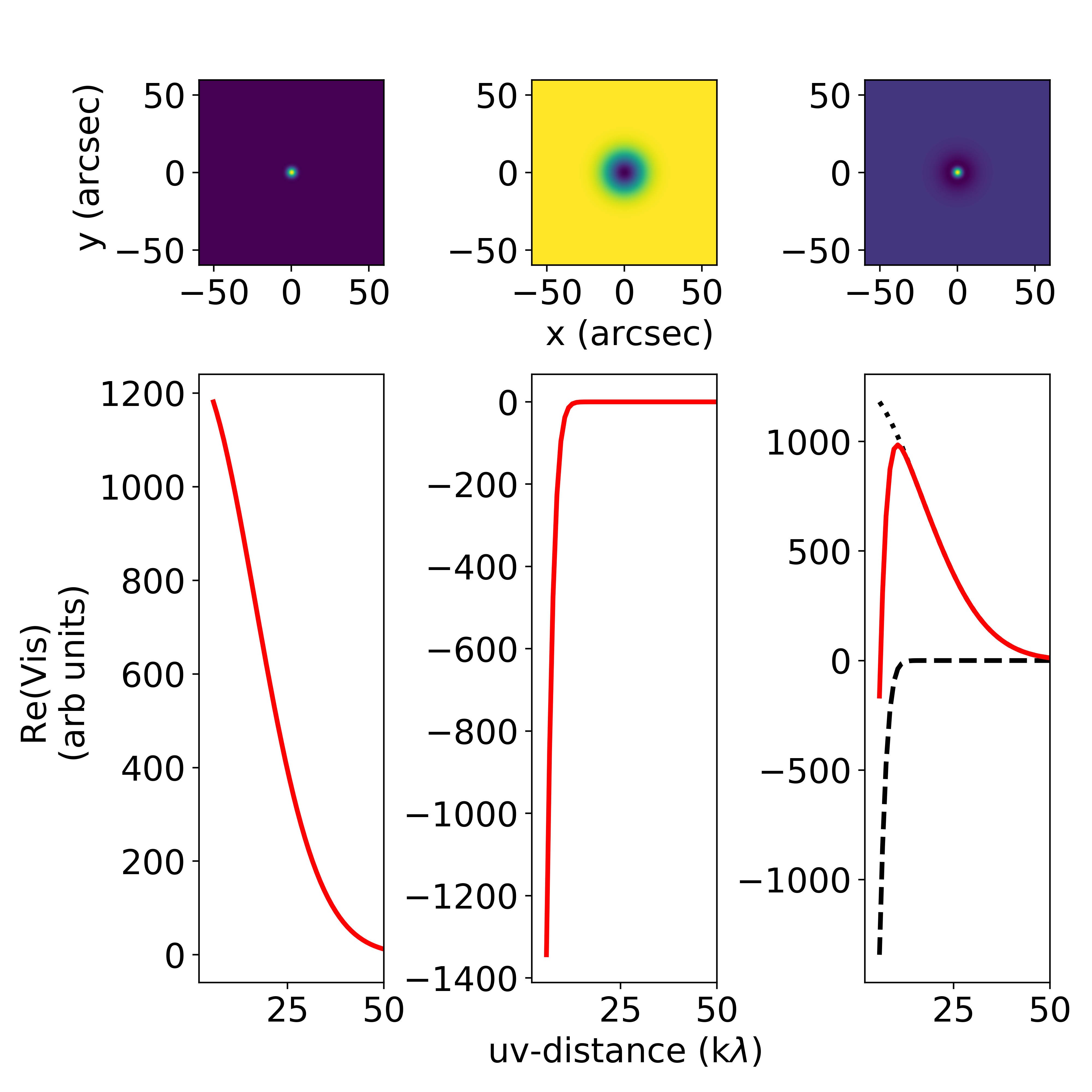}
    \caption{The top row shows three brightness profiles: a narrow, positive Gaussian signal (left), a broad, negative Gaussian signal (centre) and the superposition of the narrow, positive and broad, negative signals (right). The bottom row shows the response of an interferometer to each of these profiles: a broad, positive Gaussian (left), a narrow, negative Gaussian (centre) and the superposition of the broad, positive Gaussian (black, dotted) and narrow, negative Gaussian (black, dashed), producing a downturn in flux at the shortest baselines (right). The brightnesses of the objects are not physically motivated. They are amplified in order to demonstrate the response of an interferometer to emission on different spatial scales. All the source profiles are symmetric, so the visibilities are fully described by the real part. All sky brightnesses and corresponding Fourier transforms  in this figure are not primary beam attenuated.}
    \label{fig:SyntheticGaussians}
\end{figure}

We expect short baselines to be sensitive to any spatially extended SZ emission. The extended SZ emission is, however, \textit{resolved out} by longer baselines. The difference in the angular
scales between extended and compact emission thus allows us to infer model parameters in the $(u,v)$ plane. The principle is most simply
demonstrated by assuming a narrow Gaussian profile for the galactic emission and a broad Gaussian profile for the SZ
emission in the image plane (Fig. \ref{fig:SyntheticGaussians}). By the Fourier transform, the positive emission is a
broad positive Gaussian profile whilst the SZ emission is a narrow negative Gaussian profile in the visibility plane.
The Fourier transform is additive, therefore any SZ signal should be detectable through a downturn in flux at the shortest baselines. 

The difficulty with inferring model parameters in the visibility plane lies in finding $Vis_\mathrm{mod}(u,v)$. Calculating
$Vis_\mathrm{mod}(u,v)$ requires a number of computationally expensive operations such as  2D Fourier transforms.
These operations must be performed at every iteration of the fitting procedure. More specifically,
the process consists of modifying the image plane parameters, computing the corresponding $Vis_\mathrm{mod}(u,v)$,
running the fitting procedure and iterating. The visibility plane approach thus quickly becomes slow. There are
functions built into CASA (\textit{uvmodelfit}, for example) that allow users to fit sources, but these are designed
for CPUs and are thus inefficient for such complex fits. Moreover, \textit{uvmodelfit} has limited functionality
and flexibility. GALARIO \citep{Tazzari2017} is a new code that makes use of GPUs and multiple CPU cores to compute synthetic visibilities quickly. GALARIO allows users to choose the fitting procedure - in our case, a Bayesian MCMC sampler. This freedom is not possible inside \textit{uvmodelfit}.

Aside from speed and flexibility, GALARIO offers a further more fundamental advantage to our work, which aims to fit many emission sources in
the field, as in Fig. \ref{fig:PositiveContinuumEmission}. The  CASA  task $uvmodelfit$ does not have
the required functionality to fit these sources as it can only fit simple sky brightness distributions, such as single
Gaussians and point sources. Using CASA, therefore, one would need to fit sources sequentially. This is not optimal since all the sources in the primary beam contribute to the visibilities in ways that cannot be easily separated. 
GALARIO  allows us to fit any general sky brightness distribution to the data, including multiple emission sources. 

\subsubsection{Noise determination and weights scaling}\label{Uncertainty Scaling}
All fitting procedures require the uncertainties to be well known. Yet the accuracy of the absolute scale of the default ALMA weights is uncertain\footnote{See \href{https://casaguides.nrao.edu/index.php/DataWeightsAndCombination}{https://casaguides.nrao.edu/index.php/DataWeightsAnd Combination} for a discussion of relative and absolute weights.}. This is of little concern when working in the image plane since the image plane RMS noise level is determined by the RMS scatter of the visibility points about the unknown \textit{true} visibilities of the sky-brightness distribution and the relative scale of the weights. 
 
\begin{figure}
	\includegraphics[width=\columnwidth]{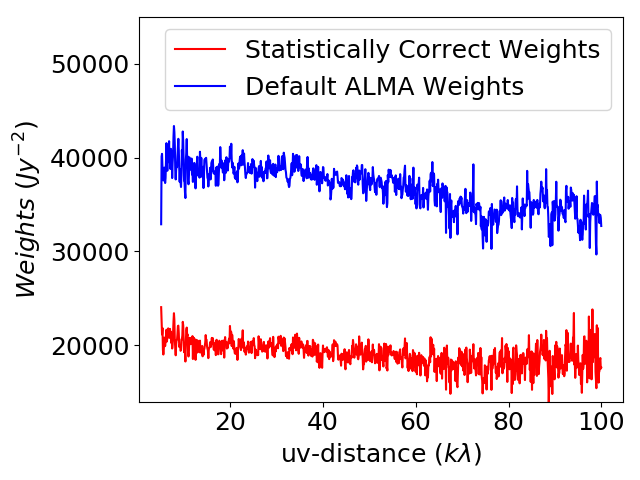}
    \caption{A comparison of the default  ALMA weights with those calculated from the statistical scatter of the visibilities. The default  ALMA weights are clearly overestimated.}
    \label{fig:WeightScale}
\end{figure}

We needed to determine the absolute scale of the weights before attempting to infer any model parameters. One expects
the standard deviation of  $Vis(\rho )$ inside a narrow bin   to correspond to the absolute uncertainty,
where $\rho  = \sqrt{u^{2}+v^{2}}$.  We define a bin as between $\rho$  and $\rho+\mathrm{d}\rho$. In practice, we require the bin width to be significantly smaller than the scale on which
$V(uvdist)$ varies in order to be considered 'narrow'. We thus calculate the  statically correct weights as the reciprocal of  the variance of the visibilities inside bins of uvdistance. We note that this procedure is roughly equivalent to that of the CASA task \textit{statwt}. We compared the statistically correct weights with the  mean of the default ALMA weights inside these bins (Fig. \ref{fig:WeightScale}). On average, the  default  ALMA weights are roughly twice as large as the statistically correct 
weights. We therefore scaled the observed weights by this difference, thus increasing the absolute scale of the
uncertainty on the individual visibility points by $\sim \sqrt{2}$.

\subsubsection{Model Parameter Estimation}\label{Model Parameter Estimation}
We used GALARIO to fit the observations with different brightness models. First, we used the  simplest
model capable of reproducing Fig. \ref{fig:PositiveContinuumEmission}, namely five point sources located at the positions of
the positive emission. The flux and positions of all the sources were free to vary. This produced  Gaussian-like
marginalised posterior distributions, but there was a clear excess in the residuals when viewed in the image
plane at
the locations of sources A, B and C. This is consistent with our initial inspection indicating that
these three sources are in fact resolved and cannot be modelled as simple point-like emission.

We therefore use a brightness model with the three resolved sources described as extended Gaussian profiles and the remaining two sources as point sources. The peak brightness and FWHM of the Gaussian sources were free to vary.  Note, we assumed symmetric Gaussian profiles (i.e. minor axis = major axis) to reduce the number of free parameters in the model. The positions of the point sources were fixed based on the results of the fit described in the previous paragraph. The 1D marginalised distributions are described by good normal probability distributions, indicative of a good fit. The width and peak brightness of the resolved sources are correlated in the 2D marginalised posterior distributions.  This is not surprising as they are only marginally resolved and the total flux of the sources are degenerate in both peak brightness and width of the Gaussian emission. The model parameters are summarised in Table \ref{table:ParameterSummary}. Note, we are able to determine angular extents of Gaussian profiles much smaller than the synthesised beam through sensitivity to the tails of the emission.

\begin{table}
 \renewcommand*{\arraystretch}{1.4}
\caption{Median parameter estimates describing the positive emission and associated errors. The superscripts and subscripts denote the 84th and 16th percentiles of the parameter 1D marginalised posterior distributions respectively.  The 1$\sigma$ uncertainty in the position parameters is $\sim$0.1 arcsec for all sources. The positions of sources D and E were not free parameters in the model. These two sources are point-like and thus  have no meaningful FWHM. The flux densities are  primary beam corrected and are reported at an effective frequency of 140 GHz.}
\centering
 \begin{tabular}{c c c c c} 
 \hline
& Flux Density & FWHM & RA & Dec \\
& ($\mu$Jy) & (arcsec) & & \\
 \hline
A & $70.6^{\mathrm{+4.5}}_{\mathrm{-4.3}}$ &  $0.40^{\mathrm{+0.22}}_{\mathrm{-0.14}}$ &  05:17:07.61 & -44.10.55.5 \\
B &  $61.2^{\mathrm{+6.4}}_{\mathrm{-4.3}}$ & $0.72^{\mathrm{+0.39}}_{\mathrm{-0.37}}$  & 05:17:07.73 & -44.10.44.0 \\
C & $387.1^{\mathrm{+9.5}}_{\mathrm{-9.2}}$ &  $0.49^{\mathrm{+0.12}}_{\mathrm{-0.12}} $ &  05:17:08.93 & -44.10.52.4 \\
D & $25.1^{\mathrm{+4.2}}_{\mathrm{-4.2}}$  & N/A  &  05:17:08.24 & -44.10.51.4 \\
E & $34.9^{\mathrm{+10.1}}_{\mathrm{-10.8}}$ & N/A & 05:17:09.19 & -44.11.00.9\\
 \hline
 \end{tabular}

\label{table:ParameterSummary}
\end{table}

\begin{figure}
	\includegraphics[width=\columnwidth]{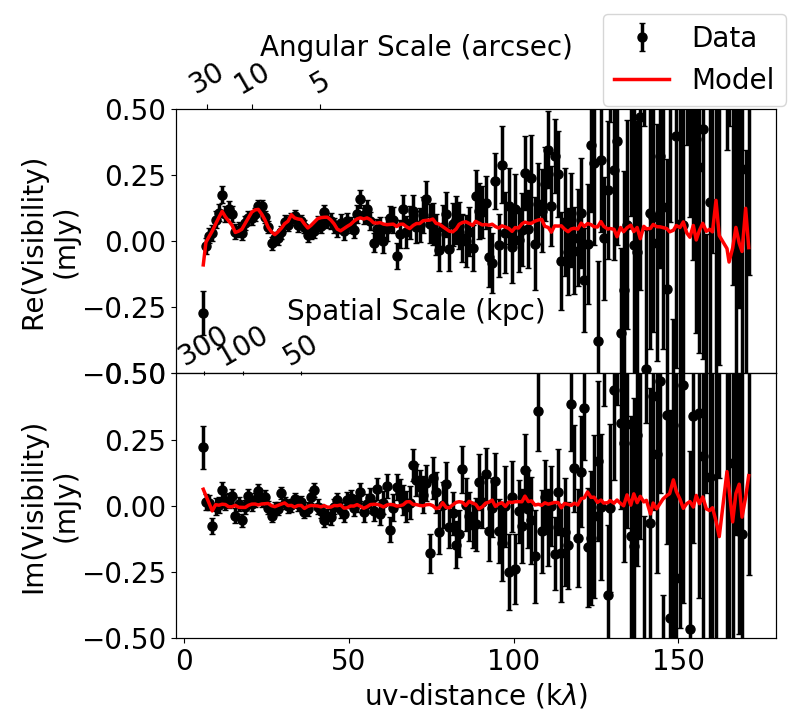}
    \caption{The visibility measurements (black points) and best fitting model visibilities (red line)
	describing the positive emission in bins of width 1~k$\lambda$. The top and bottom panels show the real 	and imaginary parts of the visibilities with error bars representing the standard error on the mean. The long baseline data points have significantly larger uncertainties due to lower
	uv-coverage. The y-axis scale is cropped to highlight oscillating visibilities caused by the positive emission in the field.}
        \label{fig:VisibilityDataModel}
    \end{figure}

We present the real and imaginary parts of $Vis_\mathrm{obs}(uvdist)$  and $Vis_\mathrm{mod}(uvdist)$  in Fig. \ref{fig:VisibilityDataModel}. Overall, the model fits the long baseline data well,  accurately following the oscillations in  $Vis_\mathrm{obs}(uvdist)$ produced by the offset positive emission, but there is a discrepancy between the model and the data at the very shortest baseline (5-6~k$\lambda$). We will examine this residual emission in Section \ref{Real Observations Visibility Plane} and quantify its significance in Section \ref{Significance}. We note that there is also some discrepancy between the model and the data in the range 80-120~k$\lambda$. This is probably due to inaccurate modelling of the positive sources' extents, particularly in modelling source C with a symmetric Gaussian profile. We examine this further in Section \ref{Image Plane}. 

The Hermitian symmetry of visibilities forces the imaginary part to average to zero in annuli of radial uvdistance. This limits the information that can be obtained from analysing  the imaginary part as a function of uvdistance. Significant non-zero imaginary parts are only possible within radial annuli that are not sampled completely. This effect can be seen in Fig. \ref{fig:VisibilityDataModel}. We observe non-zero imaginary flux at the very shortest baseline, a uvdistance bin that is not fully sampled. The imaginary component signals that the source structure is not completely symmetric, but we cannot constrain this further when examining the visibilities as a function of uvdistance. We note that the fluctuations in the imagianry part at longer baselines are also caused by uneven sampling of uvdistance annuli.

We  attempted to directly model the SZ absorption by fitting a negative, very broad Gaussian.
Unfortunately, this modelling does not provide well constrained results for
the absorption component, probably due to the fact that the SZ has a much more complex structure (see Section
\ref{Comparison with Simulations} and e.g. Fig. \ref{fig:SZband4}). Indeed, cosmological simulations
expect the S-Z signal to have a multi-shell like structure which does not have a smooth radial profile,
and certainly not a simple Gaussian profile. Any attempts to fit the visibilities describing a model SZ map from cosmological simulations presented in Section \ref{Comparison with Simulations} with a Gaussian structure using GALARIO were unsuccessful, both at low SNR, with noise at the same level as the HE0515-4414 data, and very high SNR, with SNR increased by a factor of 20. Crucially, the low SNR simulations demonstrate that the deviations from a Gaussian model are detectable at the noise level in our data. GALARIO has of course comfortably fitted other real Gaussian profiles, such as the three Gaussian positively emitting sources in this work. The inability to fit any SZ emission with GALARIO thus implies that the SZ emission has a more complex structure and does not detract from this work.

\section{Results}\label{Results}
Our model did not include an SZ component. The SZ component is expected to be present on scales much larger
than the positive emission, so its features cannot have been accounted for by any of the components in our
model. Any SZ effect stronger than ALMA's sensitivity should therefore be apparent in the residuals. The residuals were found by subtracting the model presented in Fig. \ref{fig:VisibilityDataModel} from the observed visibilities. In this  section, we examine the residuals in both the image and visibility plane. 

\subsection{Image Plane}\label{Image Plane}
\begin{figure}
	\includegraphics[width=\columnwidth]{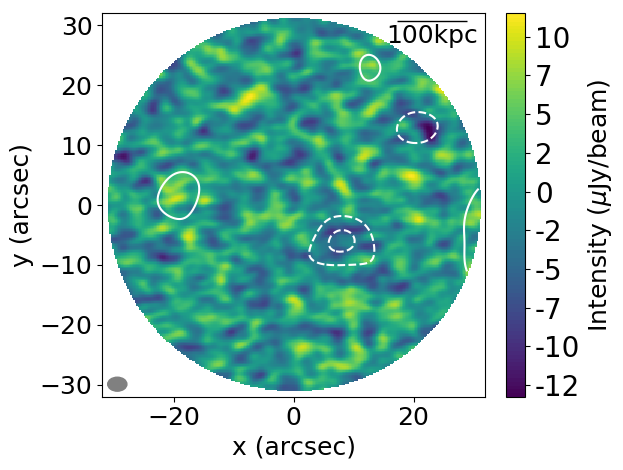}
    \caption{Dirty, primary beam attenuated image after subtracting the best fitting positive emission model.  The untapered residuals have SNR less than three. Contours are taken from a corresponding clean tapered (10 arcsec) image (in particular, Fig. \ref{fig:ImageResidualTaper}) and are shown at 2, -2, and -3$\sigma$, where the 1$\sigma$ level is at 7.8~$\mu$Jy~beam$^{-1}$.}
        \label{fig:ImageResidual}
    \end{figure}

\begin{figure}
	\includegraphics[width=\columnwidth]{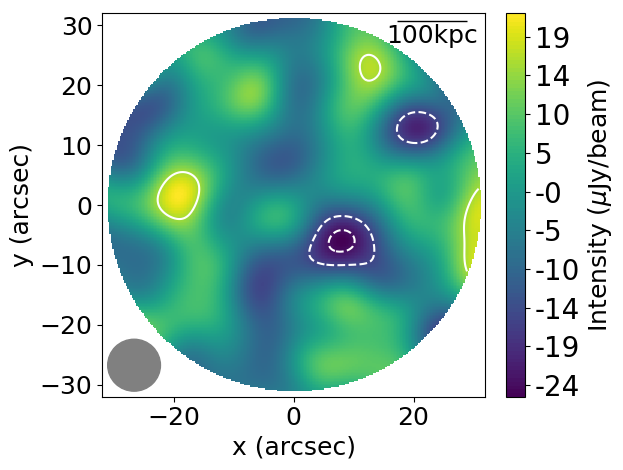}
    \caption{Clean, primary beam attenuated image of the residuals with a 10~arcsec taper applied. The 3.2 $\sigma$ hole to the southwest of the central quasar was treated as real and cleaned. Contours are are shown at 2, -2, and -3$\sigma$, where the 1$\sigma$ level is at 7.8~$\mu$Jy~beam$^{-1}$.}
        \label{fig:ImageResidualTaper}
    \end{figure}

Image plane residuals were produced inside CASA using \textit{tclean} with natural weighting. Again, this brings increased sensitivity at the expense of decreased angular resolution. Any SZ signature would be highly resolved by the array, so this trade-off is well justified. Two residual maps were produced - one without any tapering applied and a second with  a 10~arcsec  taper (Figs. \ref{fig:ImageResidual} and \ref{fig:ImageResidualTaper}). All tapering in this paper is applied using the \textit{uv-taper} parameter inside CASA's \textit{tclean}. Tapering reduces the weights of higher spatial frequencies relative to lower spatial frequencies. It is equivalent to smoothing the dirty beam and is used to filter out signals on spatial scales smaller than the beam, thereby highlighting more extended features. This amounts to artificially decreasing the angular resolution of the image. 

The untapered image, Fig. \ref{fig:ImageResidual}, looks mostly like random noise. There is some suggestion of residual positive emission, but all these features have SNR less than three. The appearance of remnant low SNR features is to be expected given the observation's high sensitivity. We decided to make no further attempt to subtract these 'sources'. We also note that there is no sign of any over-subtraction at the positions of the sources in the model (i.e. there are no holes in the brightness profiles at their locations). 

The tapered residuals  (Fig. \ref{fig:ImageResidualTaper}) are more interesting. There is a 2$\sigma$
emission feature at the location of source C. This is likely to be a consequence of modelling the source as a symmetric Gaussian. The residual emission may thus arise from any asymmetry in source C's brightness profile. We remind the reader that symmetric Gaussian profiles were used in order to reduce the number of free parameters in the model. The residual emission from source C is the likely  cause of the discrepancy between the data and the model at 80-120~k$\lambda$ , as described in Section  \ref{Model Parameter Estimation}. However, more interestingly, we note the tentative
detection of a negative ``bowl'' to the south west of the central quasar. A clean region was placed at
the location of this feature when producing Fig. \ref{fig:ImageResidualTaper}. No other region was selected
for cleaning. This is at the same location
as the dip claimed by \citet{Lacy2018} using the same data. The latter claim this dip to be the direct
detection of an SZ signal, SW of the quasar, at 3.5$\sigma$ significance. In our analysis this feature
has a slightly lower significance, 3.2$\sigma$. This "bowl" has an integrated flux of -24.5$\mu$Jy
\citep[vs -26$\mu$Jy in ][]{Lacy2018} .
The small difference with respect to \citet{Lacy2018} is likely a consequence of the combination of
three factors: 1) we have performed the source fitting and subtraction in the uv plane (hence free
from cleaning and the Fourier transform issues discussed above), while \citet{Lacy2018} has performed
the source subtraction in the image plane;
2) we have subtracted five emission sources, while \citet{Lacy2018} have subtracted only
four emission sources; 3) we have fitted three of them as resolved sources (as indeed inferred by
GALARIO), while \citet{Lacy2018} have fitted all sources as unresolved.
However, this relatively small and off-centered S-Z signal is likely to be tracing a localised heating and
not the global signal associated with the entire hot halo.
We will therefore now focus on emission
extended on larger scales based on the UV plane analysis.


\subsection{Visibility Plane}\label{Real Observations Visibility Plane}
As with much of interferometric analysis, it is more instructive to look at the residuals in the visibility
plane. Fig. \ref{fig:VisibilityDataModel}  clearly shows that the model fits the data well at long baselines.
We therefore focus on the shorter baselines ($uvdist$~<~10~k$\lambda$) in this analysis. This is where we expect to see an SZ signal. 

\begin{figure}
	\includegraphics[width=\columnwidth]{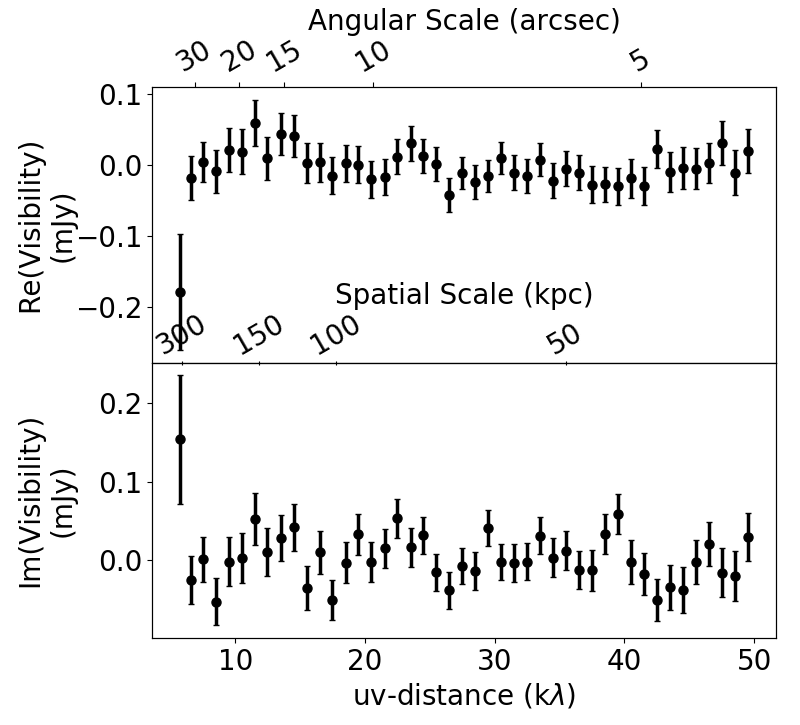}
    \caption{Residual visibilities for $uvdist$~<~50~k$\lambda$, with bin size 1~k$\lambda$. Residual flux is observed at the shortest uvdistances.}
        \label{fig:50klambdaResidualAmplitude1000}
    \end{figure}

We present the residuals in Fig. \ref{fig:50klambdaResidualAmplitude1000} in bins of 1~k$\lambda$. The residuals are consistent with a blank field at all but the
very shortest baselines, where we see flux in both the real and imaginary components.  This confirms the claim of Section \ref{Model Parameter Estimation}: the model fits
the data well at long uvdistances but the visibilities at short uvdistance cannot be fully described by the
(positive) emitting sources  in the model.

Symmetric, centred, extended, negative Gaussian signals exhibit in the visibility plane as a negative dip in the real part at short baselines and have no flux in the imaginary part. Such sources will therefore have no flux in the imaginary component when binned in annuli of uvdistance regardless of the sampling density. Asymmetric sources, however, are described by visibilities with non-zero imaginary parts. As explained in the previous section, however, the Hermitian symmetry of visibilities  forces the imaginary part to average to zero in radial bins of uvdistance when  the annulus is fully sampled. Crucially, the non-zero imaginary component does not average to zero when the annulus is not well sampled, as in the shortest baseline in Fig. \ref{fig:50klambdaResidualAmplitude1000} due to poor uv coverage.  Thus the non-zero imaginary part shows that the source is not symmetric, but the Hermitian symmetry of the visibilities makes it difficult to describe this asymmetry in detail when viewed in bins of uvdistance.  On the other hand, the limited SNR prevents us from obtaining more information about the SZ structure and geometry through full 2D fitting in the uv plane. We thus recognise that the imaginary component indicates the presence of some asymmetry, but we are unable to deduce more detailed information about the source structure and only attempt to place constrains on the angular extent and total flux of the signal.

7430 channel and time averaged data points are contained in the leftmost bin in Fig. \ref{fig:50klambdaResidualAmplitude1000} (i.e. 5-6~k$\lambda$). This corresponds to $\sim1.5\times{10}^{7}$ measured visibilities ($\sim38000$ independent samples) for each of the real and imaginary components. We plotted the uv coordinates of the visibilities inside that bin and observed over 50 uv-tracks.  Furthermore, a simple plot of baseline number vs uvdistance using CASA's \textit{Plotms} confirms that more than 20 antenna-antenna pairs have data at uvdistances less than 6~k$\lambda$.   We are thus confident that the large signal observed in Fig. \ref{fig:50klambdaResidualAmplitude1000}  is not simply the result of a single, perhaps inaccurate, antenna-antenna pair.   For comparison, the second bin from the left (i.e. 6-5~k$\lambda$)  contains 51264 averaged visibilities. This order of magnitude increase in the number of baselines explains the larger error bar in the smallest baseline bin.
 
We note that the negative signal revealed in the visibility plane is not the same as the
smaller and weak negative ``bowl'' to the SW tentatively seen in Fig \ref{fig:ImageResidualTaper}, at the
$\sim$20~$\mu$Jy level, and whose detection is also claimed by \cite{Lacy2018}. The SW "bowl" probably represents a localised heating inside the global halo heating structure.
It is therefore likely to contribute to the asymmetric/complex structure described by the imaginary part
of the residual visibilities.
The emission revealed in the visibility plane is on much larger scales, $\sim$ 30-40 arcsec ($\sim 300$~kpc),
and much deeper ($\sim 0.2$~mJy). However, being so broad and comparable with the ALMA primary beam,
this feature is not easily seen in the image plane.
    
\subsection{Significance}\label{Significance}
Quantifying the significance of the residuals depicted in the previous section is non-trivial. One usually conducts a simple ${\chi}^{2}$ analysis on the full data set, but such a test will be ineffective at computing the significance of the short baselines residuals. This is because the model fits the data well for almost all uvdistances (Fig. \ref{fig:VisibilityDataModel}). Any global  ${\chi}^{2}$ test will be dominated by the well fitted data, concealing the significance of data points with $uvdist$~<~10~k$\lambda$.  We thus want to consider the  ${\chi}^{2}$ of the short baselines alone, without the influence of the longer baseline measurements.

Such a test is complicated by the relatively low number of short baseline data points. After time and channel
averaging, we have only 18 data points (i.e. 34560 measured visibilities) with  $uvdist$<5.5~k$\lambda$.  Note, these are nine for the each of the real and imaginary
components. The two components of an interferometer are uncorrelated, so  a data set of N visibility
measurements consists of 2N independent measurements \citep{Thompson2017}.  The significance of the residuals is dependent on the number of degrees of freedom (dof) - number of averaged data points minus number of parameters, where the result is more significant in a scheme with lower dof. Our positive emission model, however, has 17 parameters, leaving dof=1.

A binned
reduced  ${\chi}^{2}$  (${\chi}_{red}^{2}$)  test
supports our previous claim: we find that the model fits the data well (${\chi}_{red}^{2}\sim$1) for all but the very
shortest baselines, where the ${\chi}_{red}^{2}$$\sim$12.
The probability of achieving these ${\chi}_{red}^{2}$ values through random statistical fluctuations can be
found using the ${\chi}^{2}$ cumulative distribution function (CDF). We find a 99.960 per cent
probability of  a ${\chi}_{red}^{2}$ smaller than the smallest baseline  bin  ${\chi}_{red}^{2}$,
corresponding to a 3.35~$\sigma$ detection.

Whilst the above analysis indicates the significance of the short baseline residuals, we have not yet
associated them with an SZ signal. We must show that these residuals could not be the result of
over/undersubtraction of the positive emission. For this, we consider $Vis_\mathrm{mod}(u,v)$. The model
visibilities, shown in Fig. \ref{fig:VisibilityDataModel}, clearly demonstrate that we would expect to see
oscillating residuals at longer baselines too and overall a strongly positive real part
if the decrement was  related to the positive emission. In fact, the first (positive) peak in Fig. \ref
{fig:VisibilityDataModel} is twice as strong as the 'decrement', and the flux at many longer baselines is also
similarly strong (and positive). We have already shown that significant residuals exist only at the shortest baselines, and we have thus ruled out positive emission as a candidate capable of explaining their production.

\section{Discussion}\label{Discussion}
In this section, we first compare our results to previous attempts at detecting an SZ signal from AGN feedback and then compare the results to those expected from cosmological simulations.

\subsection{Comparison to Previous Work}
Previous attempts at observing an SZ signal from single  galaxies have been made using single dish telescopes
with low angular resolution. \citet{Crichton2016} conducted a stacking analysis of radio-quiet quasars
observed with the Atacama Cosmology Telescope and Herschel in the redshift range 0.5-3.5 to derive the average
mm-infrared \textit{spectral energy distribution} (SED).  They model the SED as a greybody dust spectrum
together with an SZ component and claim 3-4~$\sigma$ significance for the SZ contribution, with a flux dip
of $\sim$0.1~mJy. They attribute $\sim$~70 per cent of this signal to AGN feedback. 

\citet{Soergel2017} focus on the degeneracies between the parameters describing galactic dust emission and the SZ component that arise when modelling the mm-infrared SEDs. They break this degeneracy by stacking Planck data of
100,000s galaxies at z$<$4 combined with higher frequency Akari data. They too report a SZ flux
of $\sim$0.1~mJy, but  with only 1.6~$\sigma$ significance. The dust parameters they find reduce
the significance of the \citet{Crichton2016}  result. Furthermore, they argue that the $\sim$70\% contribution of AGN feedback to
any SZ signal claimed by  \citet{Crichton2016}  is in conflict with predictions from simulations.  \citet{Hall2019}  have recently built on the work of \citet{Crichton2016} and explain the disagreement with \citet{Soergel2017}  over the significance of the AGN contribution to the total SZ signal in terms of the  Planck (5-10 arcmin) and ACT ($\sim$1 arcmin) beam sizes.  The contribution of the AGN averaged over a Planck beam may be small, but \citet{Hall2019} argue that the AGN heating is more concentrated and  contributes as much as 90 per cent of the SZ signal inside an ACT beam.

The advantage of our work with ALMA is clear. We completely bypass the issues relating to dust and SZ parameter
degeneracies by exploiting the difference in angular scales between the two sources. Our signal is more
significant and has a has a flux ($\sim$0.2~mJy) twice as large as those reported by \citet{Crichton2016} and
\citet{Soergel2017}.  We point out that this flux of 0.2~mJy has not been corrected for primary beam attenuation and flux that may have been resolved out by the array. We infer a correction factor of 2.5 using simulations in the following section, suggesting a total SZ flux of 0.5~mJy. This difference in flux can be understood as follows. Our object
is the most powerful radio quiet quasar known at z$\sim$1.7.  The characteristic luminosity of AGN has dropped
since this epoch, thus our object is one of the most powerful known quasars \citep{Barger2005}. The sample used
in \citet{Soergel2017} is composed, on average, by less extreme quasars, therefore
we expect a more modest heating and corresponding SZ signal.

As already mentioned,
\citet{Lacy2018} (the team who proposed these ALMA observations) have conducted an independent analysis of the data set used in this work with the same
intention of recovering an SZ signal. They reach the same continuum sensitivity in the image plane and
subtract positive emission in the image plane (which is potentially affected by cleaning issues and, more
generally, by the undersampled Fourier transform issues discussed above).
They do not subtract the source C, but do remove all of the other sources considered in our analysis by using a point source model for each of them,
i.e. assuming that all of the sources are unresolved. They note the inadequacy of the point source model in
the extraction of source C (we recall that our GALARIO analysis reveals that two additional sources
are resolved and leaving residuals if modelled as point sources).
On top of this we fit all sources simultaneously. However, the key
difference between our work and theirs is the $(u,v)$ plane analysis. \cite{Lacy2018} restrict their attention
to the image plane. As in Fig. \ref{fig:ImageResidualTaper}, they too report the excess of negative signal to
the south-west of the central quasar, claiming a 3.5~$\sigma$ detection and concluding  a 'direct detection'
of the SZ effect. The significance of this dip resulting from our analysis is 3.2~$\sigma$.
Wwe have been more skeptical of this image plane analysis alone and have built on it with a
more convincing study of the visibility plane. Our visibility plane analysis has revealed a signal
that is one order of magnitude stronger ($\sim0.2$~mJy)
and much broader (on $\sim$30-40 arcsec, or $\sim$300~kpc scale), close
to the ALMA primary beam. This signal is therefore difficult to reveal through a simple inspection of the residual images.

\subsection{Comparison with Simulations}\label{Comparison with Simulations}
We have used predictions from cosmological simulations for comparison with our results and also
to optimise the observing strategy of future observations.
For this, we use the results of the \textsc{fable} simulations (Feedback Acting on Baryons in Large-scale Environments) \citep{Henden2018} to simulate observations of the SZ emission with ALMA.  

\textsc{Fable}  is a set of simulations that model AGN feedback through two modes: a quasar-mode and a radio-mode.  The quasar-mode is dominant at high redshift and is associated with high Eddington ratios \citep{DiMatteo2005, Springel2005}. It assumes that a fraction of the BH's energy couples thermally and isotropically to the surrounding gas.  The radio-mode, however,
becomes prevalent at low redshift and more modest accretion rates \citep{Sijacki2007}. In this case, the AGN creates hot bubbles in
the halo. Cosmological simulations, such as Illustris, have successfully reproduced results such as the stellar
mass-halo mass relation described above by incorporating both modes in their models \citep{Torrey2014}.
\citet{Henden2018} have built on this work by introducing a duty cycle to the quasar mode. Rather than injecting
thermal energy continuously, the feedback energy is accumulated over a 25 Myr time period before being released in a single energetic event. This
overcomes problems of overcooling associated with early versions of Illustris and reduces the need for an overly
powerful radio mode, hence improving the agreement between the simulation and observations. 

\begin{figure}
	\includegraphics[width=\columnwidth]{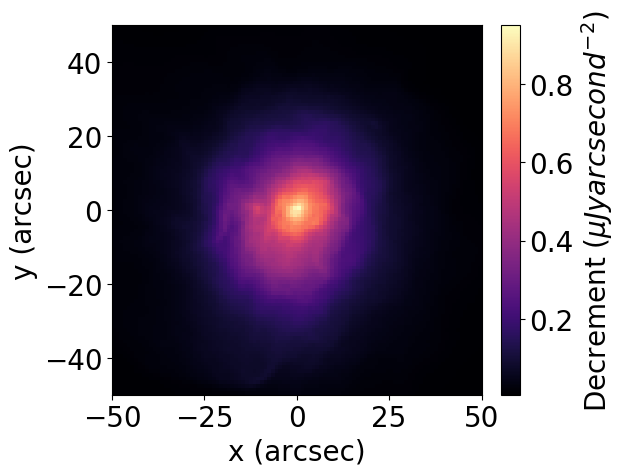}
    \caption{The SZ surface brightness associated with the simulated object described in Table \ref{table:SimulatedObject}, calculated at 140~GHz. The simulation predicts that the SZ brightness is approximately radially symmetric and does not evolve smoothly with increasing distance from the centre of the halo. This makes modelling the SZ emission difficult, as noted  in Section \ref{Model Parameter Estimation}.}
    \label{fig:SZband4}
\end{figure}

We compare the real observation discussed in this paper with simulated observations of a similar object  from the
\textsc{fable} simulations. This object was chosen from the set of ``zoom-in'' simulations of galaxy
groups and clusters described in \citep{Henden2018}. Considering only the central halo in each simulation to
ensure no contamination from low-resolution boundary particles, this object was chosen given its similarity to
HE0515-4414 in terms of black hole mass (within a factor of two; ${M}_{BH}\approx4.3\times{10}^{10}{M}_{\odot}$ \citep{Lacy2018}) and redshift. No property of the simulations other than the black hole mass and redshift were considered during this selection process. The properties of the simulated object are summarised in Table \ref{table:SimulatedObject}. The associated SZ surface brightness map as shown in Fig.  \ref{fig:SZband4} is calculated using the SZ y-parameter maps from \textsc{fable}. The map suggests a bright central core of emission surrounded by shell like structures.  This is consistent with AGN heating inflating hot buoyant bubbles, as described by radio mode feedback or through energy injected by the AGN through winds. We simulate observations of the object using the CASA task \textit{simobserve} with 10 hours of on-source time and the same spectral setup as in the real observation. The ALMA configuration is constantly changing with time as individual antennas are moved. It is thus difficult to replicate simulated observations with the exact same configuration used for observing HE0515-4414. Instead, we chose to observe the simulated SZ maps using the most compact configuration from cycle 4 - alma C40-1\footnote{See \href{https://almascience.nrao.edu/tools/casa-simulator}{https://almascience.nrao.edu/tools/casa-simulator} for configurations.}.  In practice, this array is slightly more compact than that used to observe HE0515-4414.

\begin{table}
\centering
\caption{Some basic properties of the object from the \textsc{fable} simulations that we compare to HE0515-4414. ${\dot{M}}_{BH}$ is the BH accretion rate and therefore acts as a proxy for the AGN luminosity.}
 \begin{tabular}{c c c c} 
 \hline
${M}_{BH}/{{10}^{10}M}_{\odot}$ & ${\dot{M}}_{BH}/({M}_{\odot}{year}^{-1})$ & ${M}_{500}/{{10}^{13}M}_{\odot}$ & z\\
 \hline
2.0 & 5.4 & 6.5 & 1.4\\
 \hline
 \end{tabular}

\label{table:SimulatedObject}
\end{table}


\begin{figure}
	\includegraphics[width=\columnwidth]{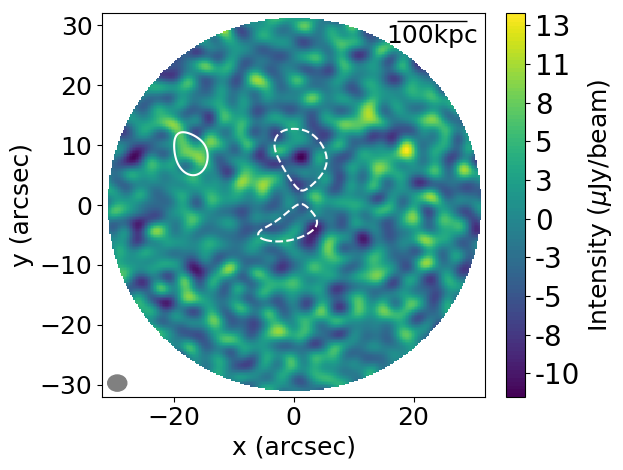}
    \caption{Dirty, primary beam attenuated image plane representation of the simulated observation for an object similar (in black hole mass and redshift) to HE0515-4414, showing an extended signal with low significance. Contours are taken from a corresponding clean tapered (10 arcsec) image and are shown at  2 and -2 $\sigma$,  where the 1$\sigma$ level is at 8.8~$\mu$Jy~beam$^{-1}$.}
    \label{fig:SZband410hrImagePlane}
\end{figure}

\begin{figure}
	\includegraphics[width=\columnwidth]{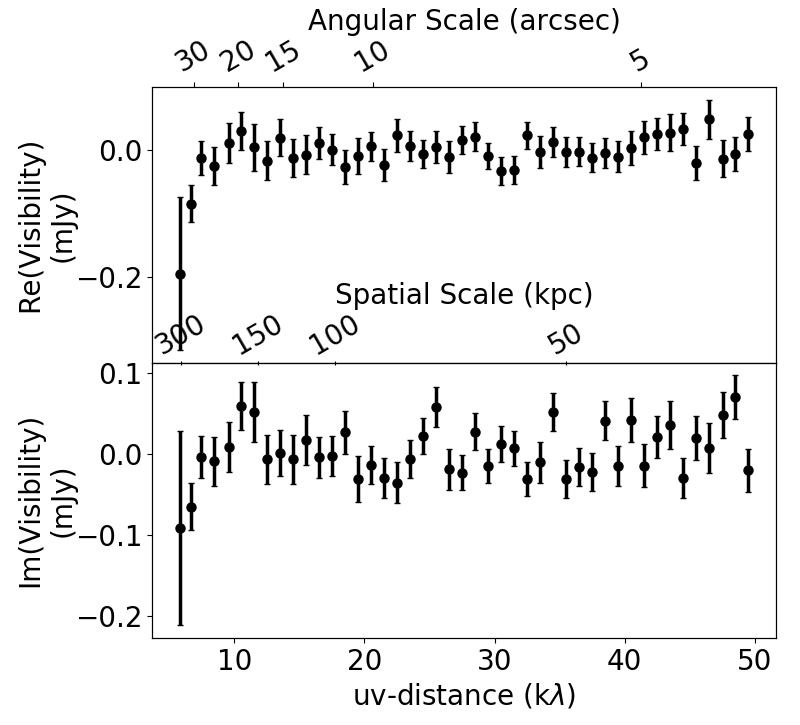}
    \caption{Visibility representation of the simulated observation for an object similar to HE0515-4414.} 
    \label{fig:SZband410hrVisibility}
\end{figure}

As with the real observation, we look at the simulated observation in both the image and visibility plane. There is weak, low significance evidence (2$\sigma$) for an extended negative SZ signal in the image plane (Fig. \ref{fig:SZband410hrImagePlane}).   \textit{Simobserve} outputs uniformly weighted visibility
measurements but sets their absolute scale arbitrarily to 1~Jy${}^{-2}$. We calculate the correct scale for the
visibility weights as in Section \ref{Uncertainty Scaling} and show the results in 
the visibility plane (Fig. \ref{fig:SZband410hrVisibility}). The data is noisy, but the real part
shows a similar form to that in Fig. \ref{fig:50klambdaResidualAmplitude1000} - i.e. we see an decrease in
flux at short baselines. The imaginary part is mostly consistent with zero. We note that, on average, the thermal energy is injected isotropically by an AGN in
\textsc{fable}, but in practice energy injection is observed to be anisotropic, both when through
winds/outflows and through radio-jets. As noted previously, the asymmetries described by the imaginary component are lost when the annuli of constant uvdistance are completely sampled. We highlight that the shortest baseline point in the real part of the actual observation and simulated visibilities are consistent within the error. However, the form of the downturn is made more clear in the simulated observations by the negative flux in the second shortest baseline.


The quasar luminosities and black hole masses of the real and simulated objects are similar. We thus expect
them to have approximately equal SZ decrement magnitudes. The analysis in this sections thus suggests that the
simulations are accurately predicting the strength of SZ signals from quasar hosts - the two measurements are consistent within the error. We argue that that
the SZ maps predicted from the \textsc{fable} simulations are consistent, in terms of flux, with that seen with a real
observation - a remarkable result given the
complex baryonic physics involved in these processes and the various potential uncertainties both in the simulations
and in the observations. 

A comparison of the flux shown in Fig. \ref {fig:SZband4} inside one primary beam with the smallest baseline point in Fig. \ref{fig:SZband410hrVisibility} allows us to estimate a correction factor for the the effects of primary beam attenuation and  resolving out of flux - i.e. flux that is distributed on scales too large to be probed by the array. Indeed, integrating over the primary beam, we measure a total flux of 0.5~mJy. We show that much of this 'missing flux' is recovered when probing shorter baselines using   Atacama Compact Array (ACA) in the next section. This suggests a total SZ flux of 0.5~mJy from HE0515-4414 after correcting for the primary beam attenuation and the flux resolved out by the array (i.e. a correction factor of 2.5). 

The small but real oscillations at longer baselines seen in Fig. \ref{fig:SZband410hrVisibility} are
noise-driven. They are not a result of any real structure in the SZ signal. We have confirmed this by examining
simulated observations of the SZ field in the absence of noise. There are
clearly no issues relating to over/under-subtraction of other sources when analysing the simulations, so
observing the oscillations here makes us fully confident in the GALARIO source subtraction for our real observation - without worrying that the small oscillations observed in Fig. \ref{fig:50klambdaResidualAmplitude1000} are due to poor source subtraction. We are confident that the oscillations cannot be attributed to the positive emission for the arguments presented in Section \ref{Significance}, and we can now confidently attribute the oscillations to noise rather than over/under-subtraction. 

The simulations also complement our understanding in the image plane. We noted in Section \ref{Image
Plane} that, whilst there may be some suggestion of an SZ decrement to the southwest of the central quasar, the image
plane case for an SZ detection is weak. There thus seems to be a disconnect between the claimed detection in the
visibility plane and the ambiguity present in the image plane. The results of Figs. \ref{fig:SZband410hrImagePlane}
and \ref{fig:SZband410hrVisibility}, however, support this scenario. We observe a short baseline negative dip in flux in  the
visibility plane, but the result in the image plane is weak. There is some excess of negative signal around the
central quasar, but it is only seen at the 2$\sigma$ level. This is simply because the SZ signal is on scales too large to be visually confirmed in the image plane.

\subsection{Optimising future observations}\label{FutureObservations}

Simulating observations of \textsc{fable} fields also allows us to explore observing strategies that would deliver
a more significant detection of the SZ signal. We consider how much scaling ('amplification') needs to be applied to the map in Fig. \ref{fig:SZband4} for a conclusive detection of
an SZ signal in band 4. Such an amplification amounts to increasing the SNR, enabling us to discriminate
features associated with real signal from those arising from noise, and is equivalent to increasing the
integration time.

\begin{figure}
	\includegraphics[width=\columnwidth]{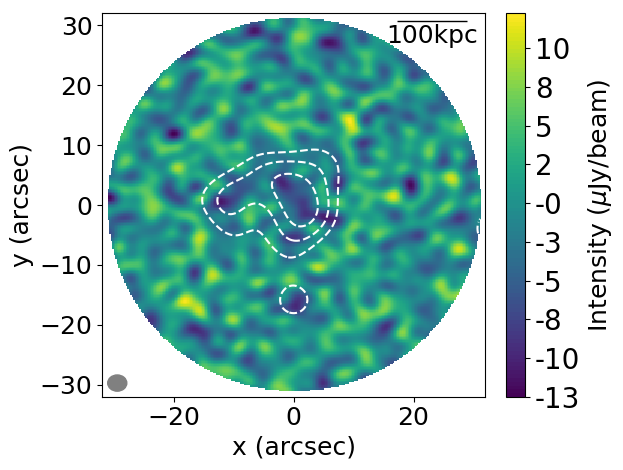}
    \caption{Dirty, primary beam attenuated image plane results for simulated observation of the SZ field in Fig. \ref{fig:SZband4} amplified by a factor of two. Contours are taken from a corresponding clean tapered (10 arcsec) image and are shown at  -2, -3 and -4 $\sigma$,  where the 1$\sigma$ level is at 8.9$\mu$Jy~beam$^{-1}$.}
    \label{fig:SZband410hrImagePlaneAmp2}
\end{figure}

\begin{figure}
	\includegraphics[width=\columnwidth]{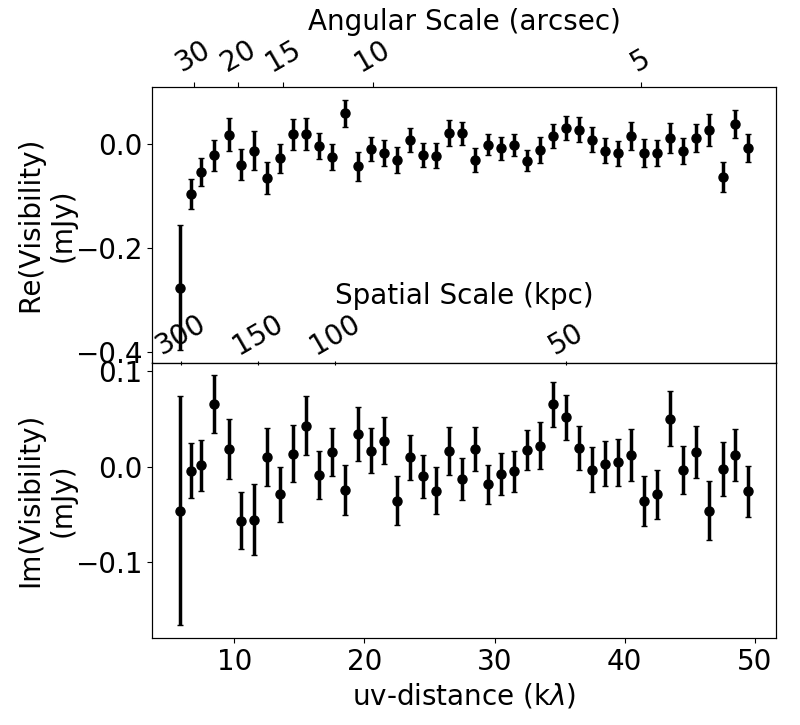}
    \caption{Visibility plane results for the doubly amplified SZ field - showing a short baseline downturn in flux and small noise-driven oscillations at longer baselines.}
    \label{fig:SZband410hrVisibilityAmp2}
\end{figure}

We begin by amplifying the SZ field used in the previous section by a factor of two (Figs.
\ref{fig:SZband410hrImagePlaneAmp2} and \ref{fig:SZband410hrVisibilityAmp2}), which is (in terms of SNR)
equivalent to increasing the integration time by a factor of four.
The signal is detected in the image plane.   The contours describing the tapered map highlight the excess of negative signal around the central quasar with 4$\sigma$ confidence. This is however only the central core of the SZ signal -
the much more extended emission predicted by the simulation is more easily seen in the visibility plane. Although the very shortest baseline measurement is highly uncertain, the significance of the downturn in flux in the real part at $uvdist$<10~k$\lambda$ is increased relative to the simulations with no amplification, as expected. We observe up to four bins outlining the form of the downturn in the real part, whilst the imaginary part is consistent with noise. By comparing Figs. \ref{fig:SZband410hrVisibility} and
\ref{fig:SZband410hrVisibilityAmp2} and assuming that the simulation predicts the  true SZ brightnesses, we have shown that increasing the SNR by a factor of two is enough to conclusively detect HE0515-4414's SZ signal
with ALMA. We therefore suggest that
future observations of this quasar in band 4 should have $\sim$40~hours of on-source time (i.e. a factor of four increase) in order
 to clearly detect HE0515-4414's SZ signature.


This work has so far focussed on observations in band 4 at $\sim$140~GHz. The observation of HE0515-4414 was proposed
at this frequency since the peak of the SZ decrement is at $\sim$130~GHz \citep{Sunyaev1972}. We argue, however, that
observations in band 3 ($\sim$100~GHz) would be better suited to making a detection. Despite the weaker SZ surface
brightness, band 3 offers two crucial advantages. Firstly, in band 3 the FOV (primary beam) of  ALMA is increased by
40 per cent. Fig. \ref{fig:SZband4} predicts an extension of the SZ signal as large as $\sim$50~arcsec. Increasing the
FOV from $\sim$40~arcsec to $\sim$60~arcsec thus dramatically increases the possibility of properly mapping the SZ
signal. The second advantage relates to the extent of the ALMA configuration in uvspace.  The absolute separation of the antennas is of course independent of the frequency, but we are exclusively concerned with the separation of the antennas in units of wavelength. In these units, the band 3 configuration is $\sim$ 40 per cent more compact than the same configuration observed with band 4. The band 3 array is thus more sensitive to extended signals. More specifically, over 10 hours of integration, the compact array used in these simulations has no visibilities with baselines under 5~k$\lambda$ when observed in band 4, yet there are almost 45000 measured visibilities recorded in the 4-5~k$\lambda$ bin in band 3. To emphasise the point, the smallest baseline bin in these simulations in band 4, 5-6~k$\lambda$, has ~2400 measured visibilities. The equivalent bin in band 3 has ~67000.  We note that the number of visibilities in the bins should not be directly compared to those discussed in Section \ref{Real Observations Visibility Plane} since the simulations were conducted using a single wide SPW rather than 4 SPW for convenience. This increased the minimum frequency observed slightly and thus reduces the sampling of the shortest baseline bins. The effect is consistent across observations in band 3 and band 4 and therefore does not detract from the claim: band 3 offers increased sensitivity to signal on larger scales.

\begin{figure}
	\includegraphics[width=\columnwidth]{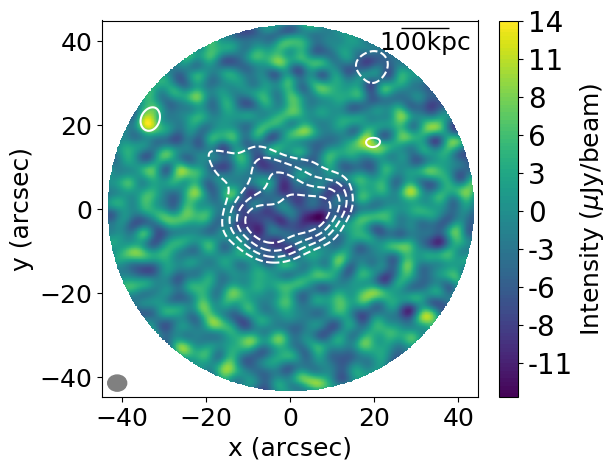}
    \caption{Clean, primary beam attenuated map for a simulated observation in band 3, without amplification.  Contours are taken from a corresponding clean tapered (10 arcsec) image and are shown at  2,-2, -3, -4 and -5 $\sigma$,  where the 1$\sigma$ level is at 5.8~$\mu$Jy~beam$^{-1}$.}
    \label{fig:SZband310hrImagePlane}
\end{figure}

\begin{figure}
	\includegraphics[width=\columnwidth]{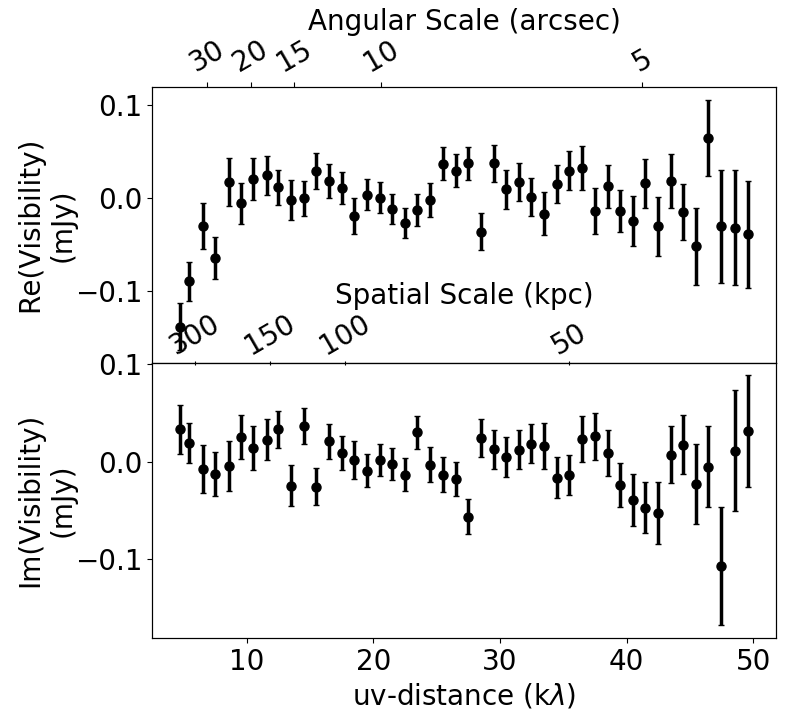}
    \caption{Visibility plane results for the SZ field observed in band 3. Note, we have removed a data point describing visibilities with baselines in the range 3-4~k$\lambda$ since it contained very few raw visibilities, two orders of magnitude fewer than the bin describing 4-5~k$\lambda$. It thus had a very large uncertainty, with 1$\sigma$ spanning almost 0.5~mJy, and provided no useful information.}
    \label{fig:SZband310hrVisibility}
\end{figure}

With the above benefits of band 3 in mind, we simulated observations of the same \textsc{fable}  field  at
100~GHz (Figs.
\ref{fig:SZband310hrImagePlane} and \ref{fig:SZband310hrVisibility}). Comparing these results with those of band 4, the SZ decrement is
apparent in both planes even without any amplification. The form of the downturn is clear in the four shortest baseline data points. As noted earlier, the intrinsic brightness of the SZ emission is greater at 140~GHz than 100~GHz. Thus, whilst we recover SZ emission with greater significance in band 3 because of the improved short baseline uvcoverage, we do not expect to recover flux with larger absolute magnitude. Indeed,  the flux shown in Fig. \ref{fig:SZband310hrVisibility} is slightly lower than that in Fig. \ref{fig:SZband410hrVisibility} when comparing bins of equivalent uvdistance. To aid this comparison, we remind the reader that the smallest bin is 5-6~k$\lambda$ in band 4 and  4-5~k$\lambda$ in band 3.  Whilst recalling consistency between the real observation
and predictions from simulations, we note the power of band 3 for use in a real observation of HE0515-4414.  We have thus demonstrated ALMAs improved sensitivity to the extended signal at lower frequencies.


We can use the results of the band 3 simulations to inform the required integration times for a real observation. The
results of Fig. \ref{fig:SZband310hrVisibility} show that the \textsc{fable} field without amplification, and thus the source
with SZ emission from HE0515-4414, is detectable with $\sim$10 hours of on-source time. We, therefore, conclude that band 3 is twice as sensitive to the extended signal as band 4, requiring only $\sim$ 25 per cent of the integration time.

\begin{figure}
	\includegraphics[width=\columnwidth]{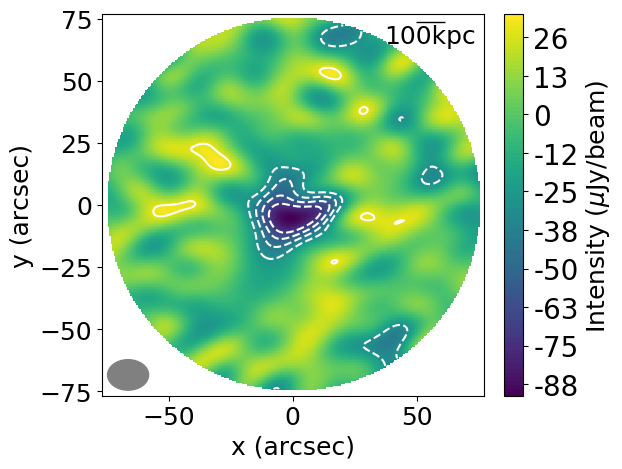}
    \caption{Clean, primary beam attenuated map for a simulated observation in band 3 using the ACA, without amplification.  Contours are drawn  2,-2, -3, -4 and -5 $\sigma$,  where the 1$\sigma$ level is at 14~$\mu$Jy~beam$^{-1}$. Note the increase in the size of the primary beam of the ACA compared to that of the 12m array.}
    \label{fig:ACASZband310hrImagePlane}
\end{figure}

\begin{figure}
	\includegraphics[width=\columnwidth]{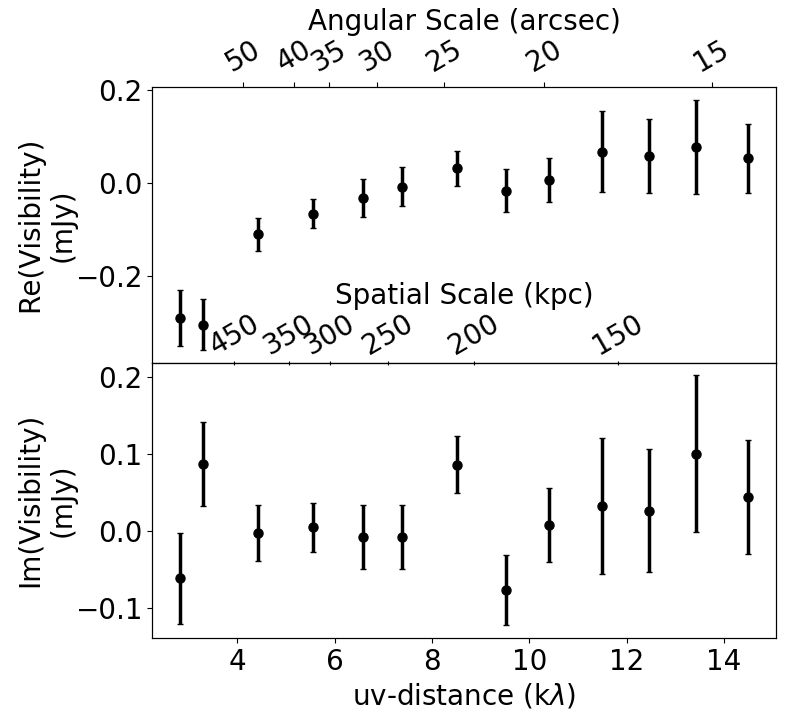}
    \caption{Visibility plane results for the SZ field observed in band 3 with ACA, enabling us to probe baselines shorter than 4~k$\lambda$.}
    \label{fig:ACASZband310hrVisibility}
\end{figure}

We have thus far been limited to observing visibilities on antenna separations greater than 12m. We need a more compact array to probe uvdistances shorter than those discussed above. The ACA, also situated on the ALMA site, consists of twelve 7m antennas capable of probing baselines nearly half the length of those probed by the 12m array. The obvious drawback of the 7m array is its reduced collecting area due to a smaller dish size and number of antennas. Much of this reduction in sensitivity can be compensated by long integration times. Observing time with ACA is typically less competitive that with ALMA. Crucially, the ACA can add valuable information about the visibility profile on very short uv distances - does the downturn increase rapidly or plateau?  We have simulated an observation of the \textsc{fable} object with 100 hours of integration in band 3 using the ACA (Figs. \ref{fig:ACASZband310hrImagePlane} and \ref{fig:ACASZband310hrVisibility}). The visibility plane confirms the benefits of the ACA.  Firstly, the results are consistent with those shown in Fig. \ref{fig:SZband310hrVisibility}, with both data sets detecting a negative dip on scales of $\sim$0.1~mJy  in the 4-5~k$\lambda$ bin.  Secondly, the two shortest baselines points, with baselines less than 4~k$\lambda$, demonstrate that the visibility profile decreases sharply. The increase in flux at short baselines confirms that the smallest baseline point in Fig. \ref{fig:SZband410hrVisibility} is missing some of the flux shown in the \textsc{fable} SZ map.  We detect the SZ decrement with 5$\sigma$ confidence in the image plane but with  low angular resolution - the ACA's large synthesised beam  is (16.5$\times$12.3~arcsec$^{2}$). Data  from ALMA and ACA should be combined to characterise the SZ emission on a range of spatial scales \citep{Kitayama2016}.


\section{Conclusion}\label{Conclusion}
We have analysed  a deep ALMA observation of the most luminous quasar at z$\sim$1.7 and used GALARIO to isolate
the SZ emission. We have considered the results of this work in the context of previous attempts and in
the context of cosmological simulations and used predictions from these simulations to demonstrate ALMAs potential in this field. Our conclusions are as follows:
\begin{itemize}
\item We find evidence for a negative dip in flux at short baselines after subtracting the positive emission from the
HE0515-4414 visibilities. The significance of the SZ detection is 3.35$\sigma$.
\item We report a SZ flux of $\sim$ 0.2~mJy ($\sim$ 0.5~mJy after applying a correction factor for primary beam attenuation and flux resolved out by the shortest baselines)  which is larger than previous estimates based on stacking of
single-dish data obtained through the stacking of 100,000s quasars. The quasar investigated in this work is more luminous than the average of the objects in previous stacking analyses, so a stronger SZ signal is expected. 
\item The imaging data also reveal a 3.2$\sigma$ negative feature to the SW of the
quasar, already claimed by \cite{Lacy2018}.  This feature is one order of magnitude
weaker than the SZ signal revealed in the visibilities and on much smaller scales $\sim$10 arcsec. This southwestern "bowl" may indicate a local region subject to additional heating.
\item The short baseline visibilities of HE0515-4414 are consistent with those of a simulated observation of a
similar object from the \textsc{fable} cosmological simulations. This demonstrates remarkable consistency
between observations and cosmological simulations. 
\item The analysis of cosmological simulations confirms the power of the analysis in visibility plane over the image plane. Low SNR
SZ emission can be seen as a negative dip in flux in the visibility plane at short baselines before a clear negative hole appears
in the image plane. Moreover, the ``hole'' seen in the image plane generally traces the central core of the SZ,
while the extended component is better detected in the visibilities.
\item We show that band 3 is more effective for detecting extended emission. This is due  both to the larger
primary beam and better short baseline coverage.  We also show the power of the ACA for probing the steep increase in SZ signal at baselines shorter than  $\sim$4~k$\lambda$.
\item Our simulated observations show that a total of $\sim$40 hours and $\sim$10 hours of on-source time are required
for a clear detection in bands 4 and 3 respectively.  Band 3 is thus twice as sensitive to SZ emission than
band 4. We argue that, combined with the current data set, a further 30 hours of on-source time observing HE0515-4414 in band 4 or 7.5 hours in band 3 would
be sufficient to make a clear detection of the SZ signal.
\end{itemize}

We have thus demonstrated ALMAs clear potential for detecting SZ emission from single galaxies. Future observations with the interferometer should enable astronomers to probe feedback physics and begin constraining models.

\section*{Acknowledgements}
This paper makes use of the following ALMA data:ADS/JAO.ALMA\#2016.1.00309.S. ALMA is a partnership of ESO(representing its member states), NSF (USA) and NINS(Japan), together with NRC (Canada), MOST and ASIAA (Taiwan), and KASI (Republic of Korea), in cooperation with the Republic of Chile. The Joint ALMA Observatory is operated by ESO, AUI/NRAO and NAOJ.
RM and SC acknowledge ERC Advanced Grant 695671 "QUENCH". SB, RM and SC acknowledge
support by the Science and Technology Facilities Council (STFC). M.T. has been supported by the DISCSIM project, grant agreement 341137 funded by the European Research Council under ERC-2013-ADG and by the UK Science and Technology research Council (STFC).




\bibliographystyle{mnras}
\bibliography{BibFiles/SZPaper,BibFiles/Books,BibFiles/FirstYearReport,BibFiles/SummerReferences}





\bsp	
\label{lastpage}
\end{document}